\documentclass[12pt]{article}
\usepackage{graphicx}
\usepackage{calc,ifthen,epsfig,miniplot,wrapfig,subfigure,rotfloat}

\input epsf
\def\hybrid{\topmargin 0pt      \oddsidemargin 0pt
        \headheight 0pt \headsep 0pt
        \voffset=-0.5cm
        \textwidth 6.25in       
        \textheight 9.5in       
        \marginparwidth 0.0in
        \parskip 5pt plus 1pt   \jot = 1.5ex}
\catcode`\@=11
\def\marginnote#1{}

\newcount\hour
\newcount\minute
\newtoks\amorpm
\hour=\time\divide\hour by60
\minute=\time{\multiply\hour by60 \global\advance\minute by-\hour}
\edef\standardtime{{\ifnum\hour<12 \global\amorpm={am}%
        \else\global\amorpm={pm}\advance\hour by-12 \fi
        \ifnum\hour=0 \hour=12 \fi
        \number\hour:\ifnum\minute<10 0\fi\number\minute\the\amorpm}}
\edef\militarytime{\number\hour:\ifnum\minute<10 0\fi\number\minute}

\def\draftlabel#1{{\@bsphack\if@filesw {\let\thepage\relax
   \xdef\@gtempa{\write\@auxout{\string
      \newlabel{#1}{{\@currentlabel}{\thepage}}}}}\@gtempa
   \if@nobreak \ifvmode\nobreak\fi\fi\fi\@esphack}
        \gdef\@eqnlabel{#1}}
\def\@eqnlabel{}
\def\@vacuum{}
\def\draftmarginnote#1{\marginpar{\raggedright\scriptsize\tt#1}}
\def\draftlabel#1{{\@bsphack\if@filesw {\let\thepage\relax
   \xdef\@gtempa{\write\@auxout{\string
      \newlabel{#1}{{\@currentlabel}{\thepage}}}}}\@gtempa
   \if@nobreak \ifvmode\nobreak\fi\fi\fi\@esphack}
        \gdef\@eqnlabel{#1}}
\def\@eqnlabel{}
\def\@vacuum{}
\def\draftmarginnote#1{\marginpar{\raggedright\scriptsize\tt#1}}

\def\draft{\oddsidemargin -.5truein
        \def\@oddfoot{\sl preliminary draft \hfil
        \rm\thepage\hfil\sl\today\quad\militarytime}
        \let\@evenfoot\@oddfoot \overfullrule 3pt
        \let\label=\draftlabel
        \let\marginnote=\draftmarginnote
   \def\@eqnnum{(\theequation)\rlap{\kern\marginparsep\tt\@eqnlabel}%
\global\let\@eqnlabel\@vacuum}  }

\def\numberbysection{\@addtoreset{equation}{section}
        \def\theequation{\thesection.\arabic{equation}}}

\def\underline#1{\relax\ifmmode\@@underline#1\else
        $\@@underline{\hbox{#1}}$\relax\fi}

\def\titlepage{\@restonecolfalse\if@twocolumn\@restonecoltrue\onecolumn
     \else \newpage \fi \thispagestyle{empty}\c@page\z@
        \def\thefootnote{\fnsymbol{footnote}} }

\def\endtitlepage{\if@restonecol\twocolumn \else  \fi
        \def\thefootnote{\arabic{footnote}}
        \setcounter{footnote}{0}}  
\relax

\hybrid

\newfont{\Bbb}{msbm10 scaled 1\@ptsize00}
\newfont{\Bbbb}{msbm7 scaled 1\@ptsize00}
\newcommand{\CC}{\mbox{\Bbb C}}

\newcommand{\DD}{\mbox{\Bbb D}}
\newcommand{\DDD}{\raise-1pt\hbox{$\mbox{\Bbbb D}$}}
\newcommand{\HH}{\mbox{\Bbb H}}
\newcommand{\HHH}{\mbox{\Bbbb H}}

\newcommand{\RR}{\mbox{\Bbb R}}

\newcommand{\UUU}{\raise-1pt\hbox{$\mbox{\Bbbb U}$}}

\newcommand{\z}{\raise-1pt\hbox{$\mbox{\Bbbb Z}$}}

\def\beq{\begin{equation}}
\def\eeq{\end{equation}}
\def\p{\partial}

\def\DD{{\sf D}}

\def\BB{{\sf B}}

\begin{document}
\begin{titlepage}

\title{Laplacian growth in the half plane}

\author{D. Vasiliev
\thanks{ITEP, Bol. Cheremushkinskaya str. 25, 117259 Moscow, Russia}
\and
A. Zabrodin
\thanks{Institute of Biochemical Physics,
Kosygina str. 4, 119991 Moscow, Russia
and ITEP, Bol. Cheremushkinskaya str. 25, 117259 Moscow, Russia}}

\date{December 2009}
\maketitle

\vspace{-6.0cm}

\begin{center}
\hfill ITEP-TH-73/09\\
\end{center}
\vspace{4.5cm}

\begin{abstract}

We investigate a version of the Laplacian growth problem
with zero surface tension in the half plane and
find families of self-similar exact solutions.

\end{abstract}

\vfill

\end{titlepage}

\section{Introduction}

In this paper we find some self-similar exact
solutions of the following version of the Laplacian
growth problem in the half plane $\HH$ introduced in \cite{Z09}.
Let $\gamma$ be a smooth non-self-intersecting
curve in $\HH$ from a point
$x_{-}\in \RR$ to a point $x_{+}\in \RR$
(we assume that $x_{-}\leq  x_{+}$) and $\BB$ be the
domain bounded by this curve and the segment $[x_- , x_+]$
of the real axis. (In \cite{Z09} such domains are called
fat slits.) Suppose the curve $\gamma =\gamma (T)$ moves with
time $T$ according to
the Darcy law:
\beq\label{gr1}
V_n (z )=\frac{1}{2}\, \p_n \phi (z )\,, \quad z \in \gamma \,.
\eeq
Here $\p_n$ is the normal
derivative at the boundary, with
the outward looking normal vector, $V_n (z )$ is the normal velocity
at the point $z \in \gamma$ and $\phi$ is a unique harmonic function
in $\HH \setminus \BB$ such that
\begin{itemize}
\item[(i)]
$\phi =0$ on $\gamma$ and on the
rays of the real axis $[-\infty , x_- ]$, $[x_+ , +\infty ]$;
\item[(ii)]
$\phi (z) = {\cal I}m \, z +o(1)$ as ${\cal I}m \, z \to +\infty$.
\end{itemize}
As is argued in \cite{Z09}, the growth process is well defined
if both angles $\alpha_{\pm}$ between
$\gamma$ and the real axis at the points $x_{\pm}$ are acute.
Then these angles
as well as the points $x_-$, $x_+$ stay fixed all the time.

Comparing this setting with the standard Laplacian growth
in the upper half plane (see, e.g., \cite{Howison}),
we see that the conditions on the harmonic
function $\phi$ are very similar: $\phi =0$
on an infinite contour from left to right
infinity and becomes ${\cal I}m \, z$
as ${\cal I}m \, z\to +\infty$. An important
difference is that in our case,
unlike in the standard one,
only a finite part of the level line
$\phi =0$ (namely, the part which lies above the
real axis) moves according to the Darcy law while the remaining part
(the rays of the real axis) is kept fixed despite the fact that the gradient of
$\phi$ is nonzero there.

It is often convenient to treat the growing
domain $\BB$ as an upper
half of the domain $\DD = \BB \cup \bar \BB$
symmetric with respect to the real
axis, with the boundary $\Gamma =\gamma \cup \bar \gamma$. Here
$\bar \BB$ is the domain in the lower half plane
obtained from $\BB$ by complex conjugation
$z \to \bar z$.
In what follows we call such domains simply
{\it symmetric}. Then one can extend the problem
(\ref{gr1}) to the whole plain ``by reflection", i.e., by saying
that $V_n (z)=\frac{1}{2}\, \p_n \phi (z )$ for $z\in \gamma$
and $V_n (z)=-\frac{1}{2}\, \p_n \phi (z )$ for $z\in \bar \gamma$,
where $\phi$ is a harmonic function
in $\CC \setminus \DD$ such that
(i) $\phi (\bar z)=-\phi (z)$;
(ii) $\phi =0$ on $\Gamma =\gamma \cup \bar \gamma$;
(iii) $\phi (z) = {\cal I}m \, z +o(1)$ as ${\cal I}m \, z \to \pm \infty$.

\section{Formulation in terms of conformal maps}

As is customary in moving boundary problems,
we reformulate the problem in terms of time-dependent
conformal maps to (or from) some fixed reference domain
from (or to) the domain $\CC \setminus \DD (T)$, where the Laplace
equation is to be solved. In our case there are two
distinguished choices of the reference domain: the upper
half plane and the exterior of the unit circle.

\begin{figure}[tb]
\epsfysize=7cm
\centerline{\epsfbox{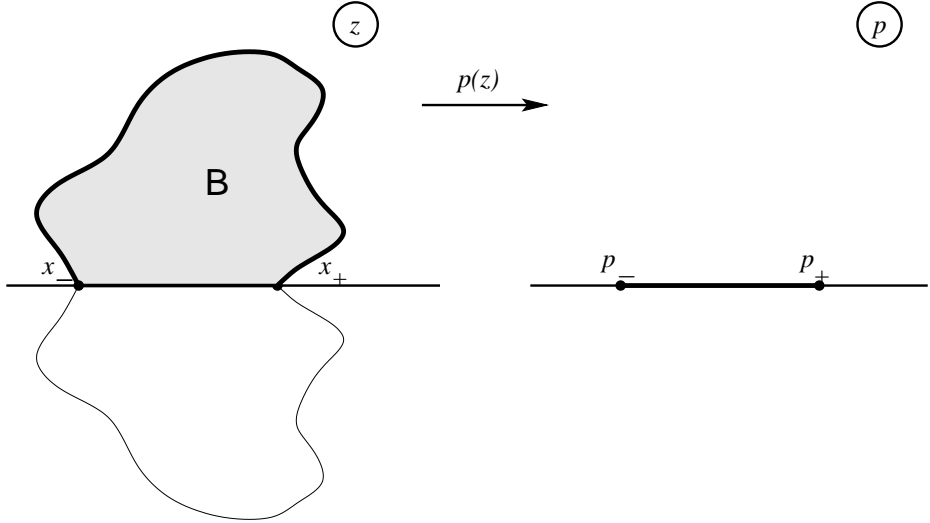}}
\caption{\sl The conformal map $p(z)$. The mirror image
of the domain ${\sf B}$
in the lower half-plane is also shown.}
\label{fig:map}
\end{figure}

\paragraph{The upper half plane.} This choice is natural
when one deals with the original formulation of the problem
in the upper half plane not extending it to the lower one
by reflection.
Let $p(z)$ be a conformal map from $\HH \setminus \BB$
(in the ``physical" $z$-plane) onto
$\HH$ (in the ``mathematical"
$p$-plane) shown schematically in Fig. \ref{fig:map}.
We normalize it by the condition that the expansion of
$p(z)$ in a Laurent
series at infinity is of the form
\beq\label{G40}
p(z)=z + \frac{u_1}{z}+\sum_{k\geq 2}u_k z^{-k}\,,
\quad |z| \to \infty
\eeq
(a ``hydrodynamic" normalization). Assuming this normalization,
the map is unique. The upper part of the boundary, $\gamma$,
is mapped to a segment of the real axis $[p_{-} , \, p_{+} ]$,
while the rays of the real axis outside $\BB$ are mapped
to the real rays $[-\infty , \, p_{-} ]$ and
$[p_{+} , \, \infty ]$. From this it follows that
the coefficients $u_k$
are all real numbers.
The first coefficient, $u_1$, is called {\it a capacity}
of $\BB$. It is known to be positive.
We also need the inverse map,
$z(p)$, which can be expanded into the inverse Laurent series
\beq\label{G41}
z(p)=p - \frac{u_1}{p} +\sum_{k=2}^{\infty}a_k p^{-k}\,,
\quad |p| \to \infty
\eeq
with real coefficients $a_k$ connected with $u_k$ by
polynomial relations.
The series
converges for large enough $|p|$.

Clearly, the harmonic function ${\cal I}m \, p(z)$ obeys
all the conditions required from the $\phi (z)$, so
$\phi (z)={\cal I}m \, p(z)$. It is easy to see that
$\p_n {\cal I}m \, p(z)= |p'(z)|$ on $\gamma$, so one can
write the Darcy law as $V_n (z) = \frac{1}{2}|p'(z)|$.

The Schwarz symmetry principle allows one to extend
this reformulation to the whole plane
as follows. The function
$z(p)$ admits an analytic continuation
to the lower half plane as $\overline{z(\bar p)}$.
The analytically continued
function performs a conformal map from the whole
complex plane with a
cut on the real axis between $p_{-}$ and $p_{+} $
onto the exterior of the symmetric domain
$\DD =\BB \cup \bar \BB$. Correspondingly, the inverse
function, $p(z)$, obeys $p(\bar z)=\overline{p(z)}$ and
$\p_n {\cal I}m \, p(z)= -|p'(z)|$ on $\bar \gamma$.
The evolution of the whole closed
contour $\Gamma = \gamma\cup \bar \gamma$ can then be written
in a unified way as
$V_n (z) = \frac{1}{2}|p'(z)|$ for any
$z\in \Gamma$.

The map $z(p)$ plays a crucial role
in the embedding of the problem into the dispersionless
KP hierarchy found in \cite{Z09} but appears to be
rather inconvenient
for constructing explicit solutions.
This is certainly related to the fact that the
symmetrically extended reference
domain in the mathematical plane
is singular (plane with a cut) and, moreover, depends
on time. There is another choice of reference domain
which seems to be less natural from the point of view of integrable
hierarchies but is more suitable for finding explicit solutions.

\paragraph{The exterior of the unit circle.}
In this case it is convenient to work with symmetrically
extended domains from the very beginning.
Let $f(w)$ be the conformal map from the exterior of the
unit circle onto the exterior of the symmetric domain $\DD$
such that $f(\infty )=\infty$ and $f'(\infty )=r>0$, so that
the Laurent expansion at $\infty$ has the form
$f(w)=rw +u_0 +O(w^{-1})$ with real coefficients.
The coefficient $r$ is called the (external) conformal
radius of the domain $\DD$.
The connection with the previously considered map
is as follows: let $w(z)$
be the inverse function, then $p(z)=r(w(z)+w^{-1}(z))+u_0$
with $r=\frac{1}{4}(p_+ - p_-)$ and $u_0=\frac{1}{2}(p_+ + p_-)$.

Let us rewrite the Darcy law as a dynamical equation for the
function $f(w)=f(w,T)$. To do that, we need
the following simple kinematical
relation which can be derived in a direct way.
Let $z(\sigma , T) =x(\sigma , T)+i y(\sigma , T)$
be any parametrization
of the contour such that $\sigma$ is a steadily increasing
function of the arc length, then
\beq \label{appA2}
V_n = \frac{d\sigma}{dl}(\p_T x\p_{\sigma} y  - \p_T y\p_{\sigma} x)
= \frac{d\sigma}{2idl}(\p_T \bar z\p_{\sigma} z  -
\p_T z\p_{\sigma} \bar z)\,,
\eeq
where $dl =|dz| =\sqrt{(dx)^2 +(dy)^2}$ is the line element along the
contour. According to our convention,
$V_n (z)$ is positive when
the contour, in a neighborhood of the
point $z$, moves to the right
of the increasing $\sigma$ direction.
Set $\sigma =-i\log w$, $z=f(w,T)$, $\bar z=f(1/w,T)$, then
the kinematical formula gives
$$
V_n (f(w))= \frac{dw}{2i|df(w)|}\Bigl (
\p_T f(1/w)\p_w f(w)-\p_T f(w)\p_w f(1/w)\Bigr )\,,
\quad |w|=1
$$
On the other hand, the Darcy law reads
$$
V_n (f(w))= \frac{1}{2}\, \frac{|dp/dw|}{|df/dw|}\,,
\quad \quad |w|=1
$$
Equating the right hand sides and using the relation
$dp/dw =r(1-w^{-2})$, we get the equation
\beq\label{G42}
w\p_w f(w)\p_T f(1/w) - w\p_w f(1/w)\p_T f(w)=
r\left |w\! -\! w^{-1}\right |\,,
\quad |w|=1
\eeq
which is the main dynamical equation of the problem
in terms of the conformal map.

It is instructive to compare it with the similar
equation for the Laplacian growth process in the whole plane
(i.e., without the condition that $\phi$ vanishes on the
the real axis) and with the same type of source at infinity.
In the latter case the evolution necessarily destroys the
reflection symmetry, so the function $f(1/w)$ should be replaced
by $\bar f(1/w)\equiv \overline{f(1/\bar w)}$ while the form
of the r.h.s. should be also changed to $-ir (w-w^{-1})$.

For finding explicit solutions of prime importance is
the case when
the function $f(w,T)$ admits an analytic continuation
across the unit circle for all $T$ in some time interval,
so that $f(w,T)$ is actually analytic not only in its
exterior but in some larger domain containing it.
However, one should take into account that
such a continuation is impossible through
the points $w=\pm 1$ which are pre-images
of the two corner points and so $f(w)$ is not
analytic there. Assuming that it is analytic everywhere else
on the unit circle, one can analytically continue equation
(\ref{G42}) as follows:
\beq\label{G43}
w\p_w f(w)\p_T f(1/w) - w\p_w f(1/w)\p_T f(w)=
\left \{ \begin{array}{l}
-ir (w-w^{-1})\,, \quad {\cal I}m \, w >0
\\ \\
\,\,\, ir (w-w^{-1})\,, \quad {\cal I}m \, w <0
\end{array} \right.
\eeq
Note that the analytic continuations to the upper and lower
half planes are different.

\section{Formulation in terms of Schwarz function}

It is known \cite{Howison}
that the standard Laplacian growth problem
in the whole plane can be integrated in terms of the
Schwarz function. For analytic contours, the Schwarz function
$S(z)$ is defined as the analytic continuation of the function
$\bar z$ away from the contour. In other words, the Schwarz function
for a curve $\Gamma$ is an analytic function $S(z)$ such that
$\bar z =S(z)$ for $z\in \Gamma$ (see
\cite{Davis} for details). If the curve depends on time,
so does its Schwarz function, $S=S(z, T)$.

In our problem the contour $\Gamma$ is not analytic because
it always contains the two corner points. However, if the curve
$\gamma$ is analytic, its Schwarz function, which we still
denote by $S(z)$,
is well defined in a strip-like neighborhood of the curve, with
the width of the strip tending to zero around the endpoints $x_{\pm}$.
We thus have $\bar z =S(z)$ for $z\in \gamma$ or, equivalently,
$\bar z =\overline{S(\bar z)}$ for $z\in \bar \gamma$.
The latter formula just means that $\bar S(z)=\overline{S(\bar z)}$
is the Schwarz function for the complex conjugate curve
$\bar \gamma$. Therefore, we can expect that
our problem can be integrated in terms of
two Schwarz functions, $S(z)$ and $\bar S(z)$,
enjoying equal rights. Below, just for brevity,
the pair $(S(z), \bar S(z))$ is referred to as
the Schwarz function of the piecewise analytic contour
$\Gamma = \gamma \cup \bar \gamma$.
In the
mathematical $w$-plane, the role of the
Schwarz function is to connect $f(w)$ and $f(1/w)$
in their common domains of analyticity:
\beq\label{M2b}
f(w)=\left \{
\begin{array}{l}\bar S(f(1/w))\,,
\quad {\cal I}m \, w >0
\\ \\
S(f(1/w))\,,
\quad {\cal I}m \, w <0
\end{array} \right.
\eeq
Note that the second equality (at ${\cal I}m \, w <0$)
is obtained from the first one by complex conjugation.

In complete analogy with the Laplacian growth problem
in the whole plane, the growth process (\ref{gr1}) is encoded
in the equation $i\p_T S  = \p_z p$ in the upper half plane.
Extending it to the lower half plane by reflection (i.e.,
complex conjugation), we have,
for all $z\in \CC \setminus \DD (T)$:
\beq\label{gr1a}
\left \{ \begin{array}{l}
i\p_T S(z,T) \, =\, \p_z p(z,T) \,, \quad {\cal}I m \, z >0
\\ \\
i\p_T \bar S(z,T) =- \p_z p(z,T) \,, \quad {\cal}I m \, z <0
\end{array} \right.
\eeq

It is convenient to rewrite the above relations in terms of
the piecewise analytic function
\beq\label{M2}
M(z)=\left \{
\begin{array}{l} \,\, i(S(z)\, - \, z), \quad {\cal}Im \, z >0
\\ \\ -i (\bar S(z)-z), \quad {\cal}I m \, z <0
\end{array}\right.
\eeq
introduced in \cite{Z09}.
By construction, it provides the analytic continuation of the function
$2|{\cal I}m \, z|$ away from the contour $\Gamma$.
Equivalently, we write $S(z)=z-iM(z)$ for ${\cal}Im \, z >0$ and
$\bar S(z)=z+iM(z)$ for ${\cal}Im \, z <0$ or, passing to the
mathematical $w$-plane,
\beq\label{M2a}
f(w)=\left \{
\begin{array}{l} f(1/w)+i M(f(1/w))\,,
\quad {\cal I}m \, w >0
\\ \\
f(1/w)-i M(f(1/w))\,,
\quad {\cal I}m \, w <0
\end{array} \right.
\eeq
In terms of the function $M=M(z,T)$ the dynamical equation
(\ref{gr1a}) acquires the form
\beq\label{TMP}
\p_T M(z,T)=\p_z p(z,T)
\eeq
Note that the function $\p_z p(z,T)$ is
holomorphic in $\CC \setminus \DD (T)$
with the expansion $\p_z p(z,T)=1+O(1/z^2)$ at infinity.

The function $M(z)$ can be uniquely decomposed as
$M(z)=M_+(z)-M_-(z)$, where $M_{+}$ is analytic in
$\DD$ and $M_{-}$ is analytic in $\CC \setminus \DD$
with zero at infinity. This decomposition is given by
the integral of Cauchy type
\beq\label{H6}
M_+ (z)=\frac{1}{2\pi i}\oint_{\p \DD}\frac{M(z')\, dz'}{z'-z}
=\frac{1}{\pi i}\oint_{\p \DD}\frac{|{\cal I}m \, z'|\, dz'}{z'-z}
\,, \quad z\in \DD
\eeq
(and the same integral for $M_-(z)$ with $z$ outside $\DD$).
The function $M_+$ is analytic everywhere in $\DD$ and, by our
assumption, can be analytically continued across the arcs $\gamma$
and $\bar \gamma$ everywhere except for their endpoints on the real
axis, where it has a singularity. One can see that
the function $M_+(z)$ is the generating function of integrals of
motion for our problem. Indeed, it is straightforward to calculate
its time derivative:
\beq\label{H7}
\p_T M_+(z)=\frac{1}{\pi i}\oint_{\p \DD}
\frac{\mbox{sign}({\cal I}m \, z')V_n (z')}{z-z'}\, |dz'|\,,
\eeq
(see \cite{Z09} for details). Plugging here
$V_n(z)=\frac{1}{2}|p'(z)|$ and recalling that
$dp(z)$ is purely real on $\Gamma$, we obtain
$$
\p_T M_+(z)=\frac{1}{2\pi i}\oint_{\p \DD}
\frac{dp(z')}{z'-z}=1
$$
where only the residue
at infinity contributes
because $p'(z)$ is holomorphic in the exterior of $\DD$
and $z$ is inside.
Therefore, we have obtained the
important equation
\beq\label{TMP1}
\p_T M_+(z,T)=1
\eeq
which allows one to construct an infinite series
of integrals of motion by expanding $M_+(z)$ into a series in $z$.

In the case of general position, when $x_- \neq x_+$,
one may expand $M_+(z)$ in a Taylor series around a point
on the real axis lying inside $\DD$ on the segment
$[x_- , x_+]$. Choosing the coordinate in such a way that
$x_- < 0 < x_+$, one can expand around the origin:
\beq\label{TMP2}
M_+(z)=T+ \sum_{k=2}^{\infty}kT_k z^{k-1}.
\eeq
Equation (\ref{TMP1}) shows that
the coefficients $T_k$, $k\geq 2$, are
conserved: $\p_T T_k =0$. They are the following harmonic moments
of the domain $\HH \setminus \BB$:
\beq\label{TMP3}
T_k = \frac{1}{\pi i k}\oint _{\DD} |{\cal I}m \, z|z^{-k}dz
= \frac{2}{\pi k} \int_{\HHH \setminus \BB}\!
{\cal I}m  \left (z^{-k}\right ) d^2 z\,, \quad k\geq 2
\eeq
One may also consider degenerate configurations with $x_-=x_+=0$.
In this case the function $M_+(z)-T$ is still conserved but
can not be represented by a Taylor series around the origin,
and so the moments (\ref{TMP3}) are ill-defined.

\section{Self-similar solutions}

Self-similar solutions are characterized by the property
\beq\label{ss1}
f(w,T)=r(T)f(w)
\eeq
which means that the time evolution is equivalent to a
dilatation in the $z$-plane. Because the time evolution
is always such that the points $x_{\pm}$ stay fixed,
self-similar solutions have chance to exist only when
$x_-=x_+=0$ which is just the degenerate case mentioned at
the end of the previous section.
Exact self-similar solutions of Laplacian growth in radial
and wedge geometries were studied in \cite{TRHC}-\cite{AMWZ07}.

\subsection{General relations}

\paragraph{Differential equation for $f(w)$.}
Substituting the self-similar ansatz (\ref{ss1})
into (\ref{G43}), we see that the variables separate:
$r(T)=T/A$ and
$$
w\p_w f(w)f(1/w) - w\p_w f(1/w)f(w)=
A\left \{ \begin{array}{l}
-i (w-w^{-1})\,, \quad {\cal I}m \, w >0
\\ \\
\,\,\, i (w-w^{-1})\,, \quad {\cal I}m \, w <0
\end{array} \right.
$$
where $A$ is a constant. This latter equation can be
treated as the Wronskian relation for a second order
differential equation of the form
$$
A_2(w)\psi '' +A_1(w)\psi ' +A_0(w)\psi =0
$$
Namely, let $\psi_1$, $\psi_2$ be two linearly independent
solutions, then their Wronsky determinant, $W=\psi _1 \psi_2'-
\psi _2 \psi_1'$ obeys the relation
$$
\p_w \log W(w) =-\, \frac{A_1(w)}{A_2(w)}
$$
In our case $W=\pm iA (1-w^{-2})$, so
$\displaystyle{
\p_w \log W = \frac{1}{w+1}+\frac{1}{w-1}-\frac{2}{w}}
$
(note that this expression is the same in the both half planes)
and the functions $f(w)$ and $f(1/w)$ are two solutions of the
ordinary differential equation
\beq\label{ss2}
\p_w^2f -\left (\frac{1}{w+1}+\frac{1}{w-1}-\frac{2}{w}\right )\p_w f +
\tilde V(w)f=0
\eeq
with yet unknown function $\tilde V(w)$, or
\beq\label{ss3}
w^2\p^2_w f - \frac{2w}{w^2 -1} \, \p_w f +
V(w)f=0
\eeq
(here $V(w)=w^2 \tilde V(w)$).

From the requirement that
$f(1/w)$ is the second solution of equation (\ref{ss3}) it follows that
$V(w^{-1})=V(w)$ and from the behavior at infinity
$f(w)=w+O(1)$ that $V(w)=O(w^{-2})$. More precisely, if
$f(w)=w+a_0+a_1 w^{-1}+ O(w^{-2})$, then a simple calculation
shows that $V(w)=2(1-a_1)/w^2 +O(w^{-3})$.
Further, from the fact that
$f(w)$ is regular outside the unit circle (except for the
simple pole at infinity) and $V(w^{-1})=V(w)$
it also follows that
$V(w)$ may have singularities only on the unit circle.
The argument is the same as for self-similar Laplacian growth
in the wedge \cite{Tu}.
Suppose, for example, that $V$ has a pole at some point
$\xi$ outside the unit circle. Then the only way
to compensate it is to impose the condition $f(\xi)=0$ which
means that the point $\xi$ actually lies on the unit circle
because the only possibility for
the point $0$ to be the image of any point $w$ such that
$|w|\geq 1$ is to be a singular boundary point in which case
its pre-image must belong to the unit circle.

\begin{figure}[tb]
\epsfysize=5.0cm
\centerline{\epsfbox{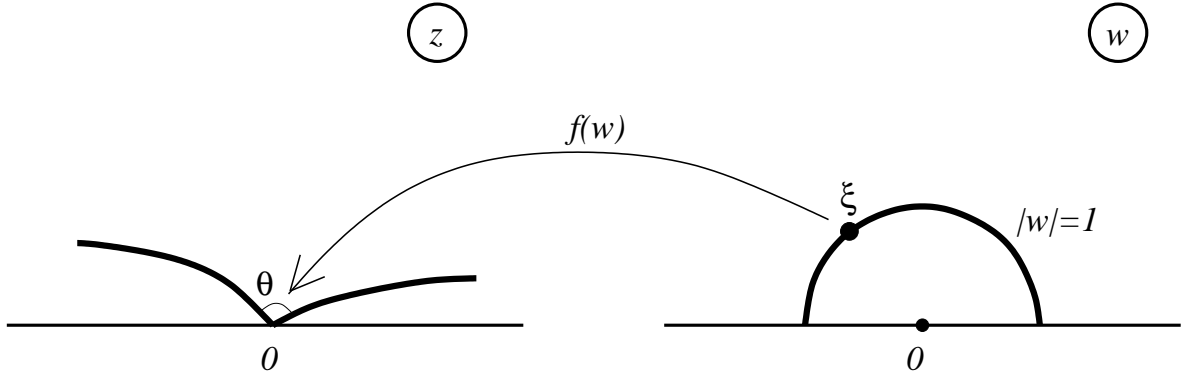}}
\caption{\sl A corner point at the origin.}
\label{fig:singular}
\end{figure}

Here we do not address the question what is the most general
type of singularities allowed for the function $V(w)$ on the
unit circle and will restrict ourselves by the case when
$V(w)$ is a rational function. Suppose the boundary has a corner
with an angle $\theta$ when $w$ moving along the
unit circle passes through a point $\xi \neq \pm 1$
(Fig. \ref{fig:singular}), then
in a vicinity of $\xi$ the conformal map has the form
$f(w)\propto (w-\xi)^{\theta /\pi}$. (We assume that
the boundary can be non-smooth only when it crosses
the origin, then $f(\xi )=0$.) Substituting it to
equation (\ref{ss3}), we get
\beq\label{ss4}
V(w)=\frac{\frac{\theta}{\pi}\left (1-\frac{\theta}{\pi}\right )
\xi^{2}}{(w-\xi)^2}\, \Bigl (1+ O(w-\xi )\Bigr ), \quad w\to \xi
\eeq
Note that the main part is the same for the angles
$\theta$ and $\pi - \theta$.
Clearly, a similar pole with the same $\theta$ is at the
complex conjugate point $\bar \xi$. Poles at $\xi =\pm 1$ should be
considered separately. In this case we get
\beq\label{ss4a}
V(w)=\frac{\frac{4\theta}{\pi}\left (1-
\frac{\theta}{\pi}\right )}{(w\mp 1)^2}\, \Bigl (1+ O(w\mp 1 )\Bigr ),
\quad w\to \pm 1
\eeq
where $\theta$ is now the angle between the real axis and
the nearest piece of the boundary curve emanating from the origin.

Let us summarize the properties of the function $V(w)$:
\begin{itemize}
\item[a)]
$V(w)$ a rational function of $w$ such that $V(w)=V(w^{-1})$;
\item[b)]
The only singularities of $V(w)$ are second order poles
on the unit circle with the principal part
given by (\ref{ss4}), (\ref{ss4a});
\item[c)]
$V(w)=O(w^{-2})$ as $w\to \infty$.
\end{itemize}
In some cases these conditions allow one to fix $V(w)$ uniquely
(see the examples below). In order to find
the conformal map $f(w)$ one should solve equation (\ref{ss3}),
choose a solution such that $f(w)=w + O(1)$ as $w\to \infty$
and check that the map is indeed conformal, i.e., that
$f'(w)\neq 0$ for all $w$ outside the unit circle.

In fact,
it is enough to check that $f(e^{i\varphi})$ makes
just one complete turn along the boundary of $\DD$
in the counterclockwise direction as $\varphi$ runs
from $0$ to $2\pi$ (the argument principle,
see, e.g., \cite{Markush}). Indeed, the number of zeros of the
analytic function $f'(w)$ outside the unit circle is given by the
integral $-\frac{1}{2\pi}\oint_{|w|=1}d\, \mbox{arg}\, f'(w)$.
(More precisely, to regularize possible singularities in
pre-images of the corner points one should consider the integral
$-\frac{1}{2\pi}\oint_{|w|=e^{\epsilon}}d\, \mbox{arg}\, f'(w)$
with a positive $\epsilon \to 0$.)
On the other hand, since $f$ is conformal in a vicinity of the
unit circle, the normal vector to the
boundary swings through the angle
$\mbox{arg}\, f'(w)$ under the map, i.e.,
$\mbox{arg}\, f'(w) =\phi (f(w))-\mbox{arg}\, w$, where
$\phi$ is the angle between the outward pointing
normal vector to the boundary of $\DD$ and the real axis.
From this it is clear that $\mbox{arg}\, f'(w)$ does not change
when $w$ makes a round trip along the unit circle and thus $f'(w)$
can not vanish outside it.

\paragraph{The form of $M_+(z)$.}
For the conformal map $p(z,T)$ self-similarity means
$p(z,T)=r(T)p(z/r(T))$ and analogously for the $M$-function:
$M(z,T)=r(T)M(z/r(T))$.
Plugging this into (\ref{TMP}), we
get $\dot r =1/A$, i.e., $r=T/A$ and
$$
M(z)-zM'(z)=Ap'(z)
$$
The function $M(z)$ here has the meaning of the $M$-function
for the domain with conformal radius $r=1$, i.e., at $T=A$.
Integrating both sides along the boundary
with the Cauchy kernel, we get the following condition
for $M_+$:
$$
M_+(z)-zM'_+(z)=A
$$
with general solution $M_+(z)=cz +A$, or
\beq\label{TMP4}
M_+(z,T)=cz +T
\eeq
where $c$ is an arbitrary real number.
It is important to
note that in the case under discussion
the domain $\DD$ consists of at least two
disconnected pieces whose boundaries intersect just
at the origin, and the constant $c$ can be different
in the different pieces.

\subsection{Examples}

\begin{figure}[tb]
\epsfysize=7cm
\centerline{\epsfbox{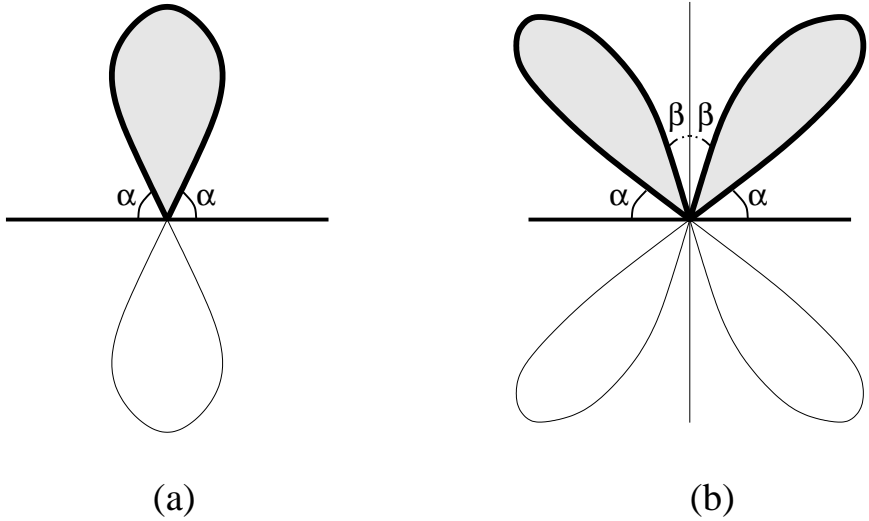}}
\caption{\sl (a) One-petal solution, (b) two-petal
solution. The mirror images
in the lower half-plane is also shown.}
\label{fig:one-two-petal}
\end{figure}

\paragraph{One-petal solution.}
The simplest example is the one-petal pattern
growing in the upper half-plane and symmetric w.r.t.
the imaginary axis (Fig. \ref{fig:one-two-petal}).
Let $\alpha$ be the angle between the tangential
lines at the origin and the real axis, then the conditions
a), b), c) above fix the function $V(w)$ uniquely:
\beq\label{e1}
V(w)=\frac{16\alpha}{\pi}\left (1-
\frac{\alpha}{\pi}\right )
\frac{w^2}{(w^2 -1)^2}
\eeq
After the substitutions $w=e^x$,
$f(w)=(\mbox{sinh}\, x) ^{1/2}\psi (x)$
equation (\ref{ss3}) takes the form
\beq\label{e2}
\psi '' +\frac{\gamma (1-\gamma )}{\mbox{sinh}^2 x} \, \psi =
\frac{1}{4}\, \psi
\eeq
where
\beq\label{e3}
\gamma = \frac{2\alpha}{\pi}-\frac{1}{2}\,,
\quad \quad -\frac{1}{2}< \gamma < \frac{1}{2}\,.
\eeq
Its solution having the required properties at infinity is
$$
\psi (x)=e^{x/2}(1-e^{-2x})^{\gamma} \, _{2}F_{1}
\left (
\begin{array}{l}
\gamma , \gamma - 1/2
\\ \phantom{aa} 1/2
\end{array}; \, e^{-2x}\right ).
$$
The hypergeometric function with these parameters can be expressed
through elementary functions:
\beq\label{elem}
_{2}F_{1} \left (
\begin{array}{l}
\gamma , \gamma - 1/2
\\ \phantom{aa} 1/2
\end{array}; \, z^2\right )=\frac{1}{2}\left [
(1+z)^{-2\gamma} +(1-z)^{-2\gamma}\right ]
\eeq
For $f(w)$ we thus obtain:
\beq\label{e4}
f(w)=\frac{1}{2}\sqrt{w^2-1} \left [
(1-w^{-1})^{\gamma}(1+w^{-1})^{1-\gamma}+
(1+w^{-1})^{\gamma}(1-w^{-1})^{1-\gamma}\right ]
\eeq
At $\gamma =0$ ($\alpha = \pi/4$) this gives the
solution $f(w)=\sqrt{w^2-1}$ mentioned in \cite{Z09}.
It is the Bernoulli lemniscate given by
\beq
\label{lemniscate}
(x^2+y^2)^2 = 2(y^2-x^2)
\eeq
in Cartesian coordinates or
$r^2=-2\cos(2\theta)$ in polar ones
(Fig. \ref{fig:1pet}, dotted line).
Using the argument principle, one can show
that $f'(w)\neq 0$ at $|w|\geq 1$
for all $\gamma$ between
$-1/2$ and $1/2$, i.e., we obtain
a continuous family of one-petal self-similar
solutions for all angles $0\leq \alpha < \pi /2$.
The examples for $\alpha=\pi /8$
and $\alpha=3\pi /8$ are shown in Fig. \ref{fig:1pet}.

\setcounter{plotFigures}{1}
\setlength{\myStandardFigureWidth}{\linewidth}
\setlength{\subSubFigPenalty}{5mm}
\includeEps{1petDashed}{One-petal solutions: $\alpha=\frac{\pi}{8}$
(dashed line), $\alpha=\frac{\pi}{4}$ (dotted line) and $\alpha=\frac{3\pi}{8}$
(solid line).}{1pet}{1}

\paragraph{One-petal solution via integral equation.}
The same results can be obtained in a different way, using
an integral equation for the conformal map $f(w)$ which follows
from the properties of the $M$-function. The function $M_+$
for the one-petal solution has only one cut which is the whole
real axis and by virtue of
(\ref{TMP4}) the jump across the cut must be a linear function.
The coefficient can be fixed by analyzing the local behavior
around the origin. Indeed,
let $e^{i\theta (z)}=dz/|dz|$ be the unit tangent vector to
the curve (represented as complex number) at a point $z$, then, by definition
of the Schwarz function,
$S'(z)=e^{-2i\theta (z)}$ for $z$ on the curve in the upper half plane
and $\bar S'(z)=e^{-2i\theta (z)}$ for $z$ on the
curve in the lower half plane. These formulas allow us to
calculate discontinuity of the function $M'$ (which obviously equals the
discontinuity of $M'_{+}$) across the
real axis.
Using (\ref{M2}) we find:
\beq
[M']:=M'_{up}-M'_{down}=\frac{1}{i}(2-S'_{up}-\bar S'_{down})
=\frac{1}{i}(2 - e^{-2i\alpha}-e^{2i\alpha})
=-4i\sin^2\alpha
\eeq
Thus in the upper half-plane
\beq
M_+=-2i\sin^2\alpha z+A,
\eeq
and in the lower one
\beq M_+=2i\sin^2\alpha z+A,
\eeq

Using
(\ref{M2b}) we can express the sum of the values of
$f(w)$ on the opposite sides of the
cut from $-1$ to $1$ in the $w$-plane. Set
$\{f(w)\}=f(w+i0)+f(w-i0)$, then
$$
\{f(w)\}=
2f(1/w)+i[M_+(f(1/w))]=(2-4\sin^2 \alpha)f(1/w)=2\cos  2\alpha \, f(1/w)
$$
where it is taken into account that
$$
[M_+(f(w))]=M_+(f(w+i0))-M_+(f(w-i0))=-4i \sin^2 \alpha f(w).
$$
We have obtained the equation
\beq\label{e9}
\{f(w)\}=2\cos  2\alpha \, f(1/w), \quad \quad w\in [-1, \, 1]
\eeq
which determines the conformal map.

It can be transformed to an integral equation
by means of the substitution
$$
g(w)=\frac{f(w)}{\sqrt{w^2-1}}
$$
which allows one to transform the mean value of the
function $f$ on the cut to the jump of the function $g$.
Note that the branch of the square root should be
chosen such that $\sqrt{w^2\! -\! 1}  =w\sqrt{1\! -\! w^{-2}}$
be an odd function outside the unit circle. The symmetry
of the petal implies that $f(-w)=-f(w)$, thus
$g(-w)=g(w)$. In terms of the function $g$
equation (\ref{e9}) acquires the form
\beq\label{e10}
[g(w)]=-2i\cos 2\alpha \, \frac{g(1/w)}{w},
\quad \quad w\in [-1, \, 1]
\eeq
The function $g$ is analytic everywhere in the
$w$-plane except for the two branch points at $w=\pm 1$,
with the jump across the cut $[-1, 1]$ being given by
(\ref{e10}). Taking into account that $g(\infty )=1$,
we can represent it by an integral of the Cauchy type
along the cut:
$$
g(w)=1-\frac{2i\cos 2\alpha}{2\pi i}\int^1_{-1}\frac{g(1/x)dx}{x(x-w)}
$$
Finally, using the fact that $g$ is an even function, we arrive at
the integral equation
\beq\label{e11}
g(w)=1-\frac{2\cos 2\alpha}{\pi}\int^1_{0}\frac{g(1/x)dx}{x^2-w^2}
\eeq
We know that $f(w)\propto
(w-1)^{\frac{2\alpha}{\pi}}$ in a vicinity of  $w=1$,
so we need a solution such that
$g(w)\propto (w-1)^{\frac{2\alpha}{\pi}-\frac12}$ near $w=1$,
and similarly for a vicinity of $w=-1$.
Note that the exponent
is the same $\gamma$ as in (\ref{e3}).
Consider first the case of positive $\gamma$.
After a further
substitution
$
g(w)=(1-w^{-2})^\gamma G(w^{-2}),
$
using the fact that
$g(\pm1)=0$ (valid for positive $\gamma$),
we arrive at the equation
\beq\label{e12}
(1-x)^{\gamma-1}G(x)=\frac{\sin
\pi\gamma}{\pi}\int^1_0\frac{t^{-1/2}(1-t)^{\gamma-1}G(t)dt}{1-xt}.
\eeq
Comparing it with equation (42) from \cite{AMWZ07}, we can immediately
write down the solution:
$$
G(t)=\, _2 F_1\left(\begin{array}{c}\gamma,\gamma-1/2\\{1/2}\end{array};t\right),
$$
which is the same function as in (\ref{elem}). The map
$f(w)$ is then given by the explicit formula (\ref{e4}).
A direct substitution shows that the function $g(w)=f(w)/\sqrt{w^2-1}$
obeys the integral equation (\ref{e11}) for both positive and
negative $\gamma$, $|\gamma |<1/2$.

\paragraph{Two-petal solution.}
Consider now a two-petal pattern growing in the upper
half-plane and symmetric w.r.t.
the imaginary axis (Fig. \ref{fig:one-two-petal}).
Let the angle between the two petals
be $2\beta$. Clearly, the angles satisfy the condition
$0<\beta < \pi /2 -\alpha$.
Now $V(w)$ has $4$ double poles.
Two of them are at $w=\pm 1$. The symmetry implies
that the other two poles are at
$w=\pm i$.  The function $V(w)$ is again uniquely determined
by the conditions a), b), c):
\beq\label{e5}
V(w)=\frac{16\alpha}{\pi}\left (1-
\frac{\alpha}{\pi}\right )
\frac{w^2}{(w^2 -1)^2}-
\frac{8\beta}{\pi}\left (1-
\frac{2\beta}{\pi}\right )
\frac{w^2}{(w^2 +1)^2}
\eeq
Note that the same function corresponds to the
angle $\pi / 2 -\beta$.
The same substitution as above,
$w=e^x$,
$f(w)=(\mbox{sinh}\, x) ^{1/2}\psi (x)$, brings
equation (\ref{ss3}) to the form
\beq\label{e6}
\psi '' +\frac{\gamma (1-\gamma )}{\mbox{sinh}^2 x} \, \psi
-\frac{\delta (1-\delta )}{\mbox{cosh}^2 x} \, \psi
=\frac{1}{4}\, \psi
\eeq
where $\gamma$ is as in (\ref{e3}) and $\delta = 2\beta /\pi$.
A further change of variables,
$$
\psi = 2(\mbox{sinh} \, x)^{\gamma}
(\mbox{cosh} \, x)^{\frac{1}{2}-\gamma}F(1/\mbox{cosh}^2  x)\,,
\quad t=\frac{1}{\mbox{cosh}^2  x}
$$
puts the equation in the canonical hypergeometric form
\beq\label{e13}
t(1-t)F_{tt}+\left (\frac{1}{2}-(1+\gamma )t\right )F_t -
\frac{1}{16}\Bigl ((2\gamma \! +\! 2\delta \! -\! 1)
(2\gamma \! -\! 2\delta \! +\! 1)\Bigr )F=0.
\eeq
Two linear independent solutions are
$$
F^{(1)}(t)= \,
_{2}F_{1} \left (
\begin{array}{l}
(2\gamma \! +\! 2\delta \! -\! 1)/4 , \,
(2\gamma \! -\! 2\delta \! +\! 1)/4
\\ \\ \phantom{aaaaaa} 1/2
\end{array}; \, t \right )
$$
and
$$
F^{(2)}(t)= \, t^{1/2}\,
_{2}F_{1} \left (
\begin{array}{l}
(2\gamma \! +\! 2\delta \! +\! 1)/4 , \,
(2\gamma \! -\! 2\delta \! +\! 3)/4
\\ \\ \phantom{aaaaaa} 3/2
\end{array}; \, t \right )
$$
For the function $\psi$ the first solution gives
$$
\psi (x)=(\mbox{sinh}\, x)^{\gamma}
(\mbox{cosh}\, x)^{\frac{1}{2}-\gamma}\,
_{2}F_{1} \left (
\begin{array}{l}
(2\gamma \! +\! 2\delta \! -\! 1)/4 , \,
(2\gamma \! -\! 2\delta \! +\! 1)/4
\\ \\ \phantom{aaaaaa} 1/2
\end{array}; \, \mbox{cosh}^{-2}x \right )
$$
Passing to the function $f(w)$ and to the original
angles $\alpha$, $\beta$, we have:
\beq\label{e7}
f(w)=w(1-w^{-2})^{\frac{2\alpha}{\pi}}
(1+w^{-2})^{1-\frac{2\alpha}{\pi}}\,
_{2}F_{1} \left (
\begin{array}{l}
\displaystyle{\frac{\alpha \! +\! \beta}{\pi} -1/2,\,\,
\frac{\alpha \! -\!  \beta}{\pi}}
\\ \\ \phantom{aaaaa} 1/2
\end{array}; \, \frac{4}{(w+w^{-1})^2}\right )
\eeq
The symmetry implies that the conformal map must be an odd function
of $w$. The function given by (\ref{e7}) is indeed odd while
the second solution, $F^{(2)}$, leads to an even function of $w$.
Using the argument principle, one can see that for
$0<\beta < \alpha <\pi /2$ the derivative $f'(w)$ has no zeros
in the exterior of the unit circle.
Therefore, $f(w)$ given by (\ref{e7}) is the conformal map
for the two-petal solution.

Note that this function looks somewhat simpler in the
variable $p=w+w^{-1}$ living in the upper half plane:
\beq\label{e8}
z(p)=p(1-4p^{-2})^{\alpha /\pi}\,
_{2}F_{1} \left (
\begin{array}{l}
\displaystyle{\frac{\alpha \! +\! \beta}{\pi} -1/2,\,\,
\frac{\alpha \! -\!  \beta}{\pi}}
\\ \\ \phantom{aaaaa} 1/2
\end{array}; \, 4p^{-2}\right )
\eeq
The analytic continuation
of the function (\ref{e8}) to the region $|p|<2$ reads:
\beq\label{e8a}
\begin{array}{r}
\displaystyle{z(p)=2(1\! -\! p^2/4)^{\alpha /\pi}
\left [ie^{-i\beta}\frac{\Gamma (\frac{1}{2})
\Gamma (\frac{1}{2} -\frac{2\beta}{\pi})}{\Gamma (
\frac{\alpha -\beta}{\pi})\Gamma (1\! -\!
\frac{\alpha +\beta}{\pi})}\left (\frac{p}{2}\right )^{2\beta /\pi}\!\!
_{2}F_{1} \left (
\begin{array}{l}
\frac{\alpha \! +\! \beta}{\pi} \! -\! \frac{1}{2},\,\,
\frac{\alpha \! +\!  \beta}{\pi}
\\ \\ \phantom{aa} \frac{2\beta}{\pi}\! +\! \frac{1}{2}
\end{array}; \, \frac{p^{2}}{4}\right )\right. }
\\ \\
\displaystyle{\left.
+ \, e^{i\beta}\frac{\Gamma (\frac{1}{2})
\Gamma (\frac{2\beta}{\pi}-\frac{1}{2})}{\Gamma (
\frac{\alpha +\beta}{\pi}-\frac{1}{2} )\Gamma (\frac{1}{2}\! -\!
\frac{\alpha -\beta}{\pi})}\left (\frac{p}{2}\right )^{1-2\beta /\pi}\!\!
_{2}F_{1} \left (
\begin{array}{l}
\frac{\alpha \! -\! \beta}{\pi} \! +\! \frac{1}{2},\,\,
\frac{\alpha \! -\!  \beta}{\pi}
\\ \\ \phantom{aa} \frac{3}{2}\! - \! \frac{2\beta}{\pi}
\end{array}; \, \frac{p^{2}}{4}\right )\right ] \phantom{aa}}
\end{array}
\eeq
The boundary of the two-petal pattern in the upper half plane
is obtained as $z(p+i0)$ for $-2 <p<2$.

In the case $\beta =\alpha$ the hypergeometric function in (\ref{e8})
becomes trivial (equal to $1$). The corresponding
conformal map appears to be a map to a slit domain: the function
$z(p)=p(1-4p^{-2})^{\alpha /\pi}$ takes the upper half plane
to the upper half plane cut along two straight segments emanating from $0$
to the NE and NW quadrants at the
angle $\alpha$ to the real axis. In fact this
means that $\beta =\pi /2 - \alpha$ rather than $\alpha$ but
as it was already mentioned, the function (\ref{e5}) is the same
in both cases.
\begin{arrangedFigure}{1}{2}{2pete}{Two-petal solutions}
\subFig[$\alpha=\frac\pi8,~\beta=\frac\pi{16}$]{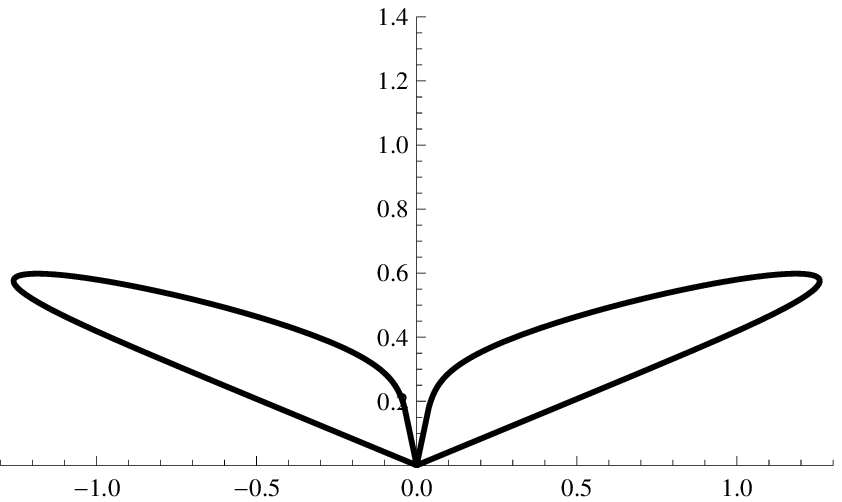}%
\subFig[$\alpha=\frac\pi8,~\beta=\frac\pi{8}$]{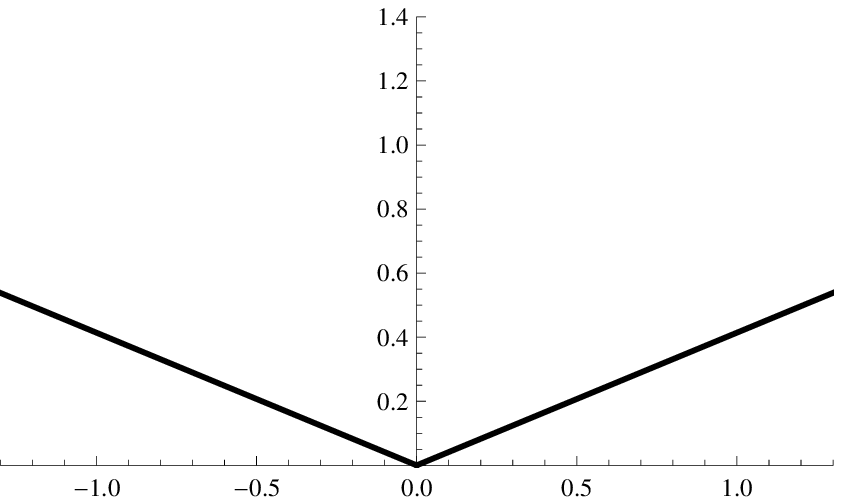}%
\subFig[$\alpha=\frac\pi4,~\beta=\frac\pi{16}$]{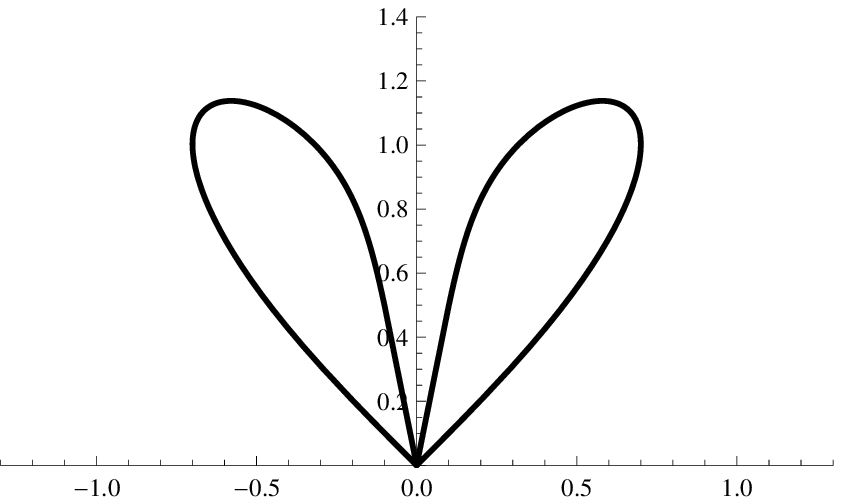}%
\subFig[$\alpha=\frac{5\pi}{16},~\beta=\frac{\pi}{16}$]{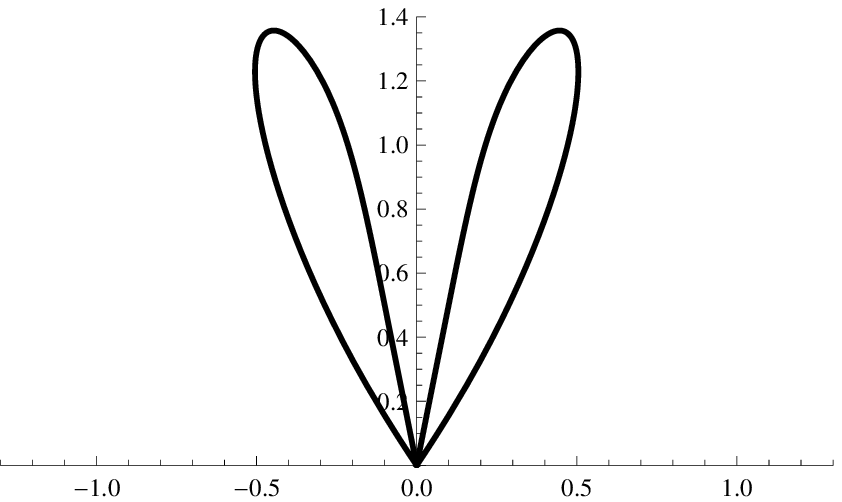}%
\end{arrangedFigure}

Some typical two-petal solutions are
shown in Fig. \ref{fig:2pete}.
The following cases are to be distinguished:
\begin{itemize}
\item[A)] $\alpha\in(0,\frac\pi4)$
\begin{itemize}
\item[i)] $\beta\in(0,\alpha)$ -- the petals are shown in Fig. \ref{fig:2pete} (a),
they become very thin as $\beta$ approaches $\alpha$ from below,
\item[ii)] $\beta=\alpha$ -- the petals degenerate to segments
(Fig. \ref{fig:2pete} (b)),
\item[iii)] $\beta\in(\alpha,\frac\pi2-\alpha)$ -- the solution
does not exist because the map is not conformal;
\end{itemize}
\item[B)] $\alpha=\frac\pi4$, $\beta\in(0,\frac\pi4 )$ --
the function $f(w)$ can be expressed through
elementary functions:
\beq
f(w)=\left(w+w^{-1}\right) \left(1-\frac{4}{\left(w+w^{-1}\right)^2}\right)^{1/4}
\cos\left[2 \left(\frac{1}{4}-\frac{\beta }{\pi }\right)
\arcsin\left[\frac{2}{w+w^{-1}}\right]\right],
\eeq
a typical pattern
is shown in (Fig. \ref{fig:2pete} (c)), the
solution degenerates at $\beta=\frac\pi4$.
\item[C)]
 $\alpha\in(\frac\pi4,\frac\pi2)$
\begin{itemize}
\item[i)] $\beta\in(\alpha,\frac\pi2)$ --
the petals are shown in Fig. \ref{fig:2pete} (d),
\item[ii)] $\beta=\frac\pi2-\alpha$ -- the petals degenerate to segments,
\item[iii)] $\beta\in(\frac\pi2-\alpha,\alpha)$ -- the solution
does not exist because the map is not conformal.
\end{itemize}
\end{itemize}
Let us also note that the shape of two-petal solutions
at small $\beta$ becomes close to one-petal solutions
with the
same $\alpha$ (Fig. \ref{fig:1pet2pet}), as it could be expected
from the differential equations.

\includeEps{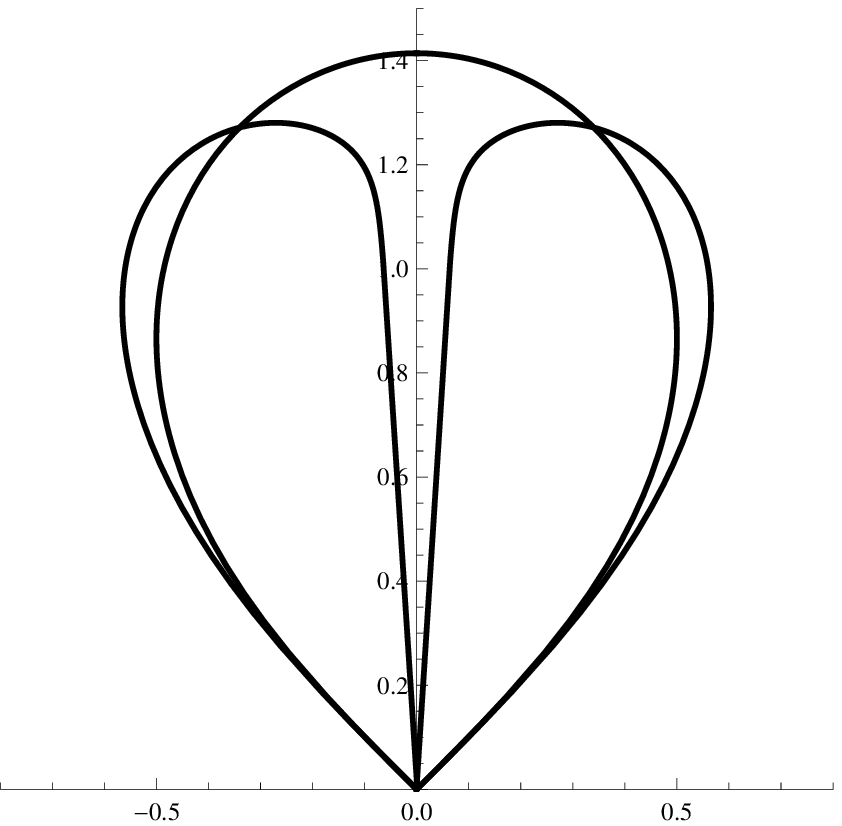}{One-petal solution at $\alpha=\frac{\pi}{4}$
and two-petal solution at $\alpha=\frac{\pi}{4}$ and $\beta=
\frac{\pi}{50}$.}{1pet2pet}{1}

\newpage

\section*{Acknowledgments}

We thank D.Khavinson, M.Mineev-Weinstein and
P.Wi\-eg\-mann for discussions. This
work was supported in part by RFBR grant 08-02-00287, by grant for
support of scientific schools NSh-3035.2008.2 and by
Federal Agency for Science and
Innovations of Russian Federation under contract 02.740.11.5029.
The work was also supported in
part by joint grants 09-02-90493-Ukr, 09-02-93105-CNRSL (D.V.)
and 09-01-92437-CEa (A.Z.).

\end{document}